\documentclass[pra,twocolumn]{revtex4}
\usepackage{amssymb}
\usepackage{amsfonts}
\usepackage{amsmath}
\usepackage{graphicx}
\usepackage{color}

\begin{document}

\title{Quantum non-Markovian environment-to-system backflows of information:
non-operational vs. operational approaches}
\author{Adri\'{a}n A. Budini}
\affiliation{Consejo Nacional de Investigaciones Cient\'{\i}ficas y T\'{e}cnicas
(CONICET), Centro At\'{o}mico Bariloche, Avenida E. Bustillo Km 9.5, (8400)
Bariloche, Argentina, and Universidad Tecnol\'{o}gica Nacional (UTN-FRBA),
Fanny Newbery 111, (8400) Bariloche, Argentina}
\date{\today }

\begin{abstract}
Quantum memory effects can be qualitatively understood as a consequence of
an environment-to-system backflow of information. Here, we analyze and
compare how this concept is interpreted and implemented in different
approaches to quantum non-Markovianity. We study a non-operational approach,
defined by the distinguishability between two system states characterized by
different initial conditions, and an operational approach, which is defined
by the correlation between different outcomes associated to successive
measurement processes performed over the system of interest. The
differences, limitations, and vantages of each approach are characterized in
detail by considering diverse system-environment models and dynamics. 
As a specific example, we study a non-Markovian depolarizing
map induced by the interaction of the system of interest with an environment
characterized by incoherent and coherent self-dynamics.
\end{abstract}

\maketitle

\section{Introduction}

The time-evolution of both classical and quantum systems may develop memory
effects \cite{vanKampen,breuerbook,vega,wiseman}. Nevertheless, the
characterization and definition of these effects is quite different in both
regimes \cite{BreuerReview,plenioReview}. As is well known, in a classical
regime memory effects can be rigorously defined in a probabilistic approach.
The independence or dependence of conditional probabilities on the previous
system history define respectively the (memoryless) Markovian and
non-Markovian regimes \cite{vanKampen}.

In a quantum regime one is immediately confronted with an extra aspect. In
fact, the state of a quantum system (and consequently its history) can only
be determined by performing a measurement process, which intrinsically
implies a perturbation to its (originally unperturbed) dynamics. Therefore,
the definition of memory effects and quantum non-Markovianity can be tackled
from two intrinsically different approaches. In \textit{non-operational
approaches}, memory effects are defined by taking solely into account the
properties of the unperturbed open system dynamics (its propagator). In 
\textit{operational approaches}, memory effects are defined by the
statistical properties of different outcomes associated to system
measurement processes and transformations (like unitary ones).

A wide variety of measures and memory witness has been utilized in the
context of non-operational approaches (see reviews \cite%
{BreuerReview,plenioReview}). The first proposals correspond to deviations
of the system propagator from divisibility \cite{divisible}\ and a
non-monotonous behavior of the trace distance\ (TD) between two distinct
system states \cite{BreuerFirst}. In this context, memory effects were
associated to an \textit{environment-to-system backflow of information}:
information stored in the initial system state is transferred to the
environmental degrees of freedom; their influence on the system at later
times implies a backflow of information that leads to memory effects. In
spite of this clear and well-motivated interpretation \cite%
{EnergyBackFLow,Energy,HeatBackFLow}, the precise assessment of this concept
is still under debate \cite%
{megier,maximal,petruccione,amato,santis,acin,horo,EntroBack}.

The basic idea of operational approaches is to appeal to the standard
definition of memory effects in terms of probabilities \cite{vanKampen}.
Hence, the (quantum) system must be subjected to a set of measurement
processes such that their statistical properties determine the presence or
absence of memory effects \cite{modi,budiniCPF}. The study and understanding
of this approach has been performed in the recent literature \cite%
{pollock,pollockInfluence,bonifacio,han,ban,rio}, including alternative
definitions and analysis of information flows \cite{goan,BIF}.

The main goal of this paper is to analyze and to compare how the concept of
environment-to-system backflow of information is interpreted and implemented
in operational and non-operational approaches. As a non-operational memory
witness we take the TD between two different system initial states \cite%
{BreuerFirst}, taking also into account the bounds on its revival behavior
that has been characterized recently \cite{EntroBack}. As an operational
memory witness we consider a conditional past-future (CPF) correlation \cite%
{budiniCPF}, both in a deterministic and random schemes \cite{BIF}. The
comparison is performed by considering different system-environment models
and analyzing in each case the information flows from the two perspectives.
We consider statistical mixtures of Markovian system evolutions, system
coupled to incoherent \cite{maximal}\ and coherent casual bystander
environments \cite{casual}, which are characterized by a self-dynamics that
is independent of the system degrees of freedom. In addition, we consider
(standard) unitary system-environment models \cite{breuerbook}. 
As a specific model, we study a depolarizing map induced by the interaction
of a system with a finite set of incoherent degrees of freedom. In this
regime, as well as in a quantum coherent one, we explain how and why both
approaches lead to different notions of quantum non-Markovianity and
environment-to-system backflows of information. 

The paper is outlined as follows. In Sec.~II we review the definition and
main properties of the considered non-operational \cite%
{BreuerFirst,EntroBack}\ and operational \cite{budiniCPF,BIF} approaches. In
Sec. III we study both approaches by considering different
system-environment models. In Sec.~IV we study the depolarizing map. In Sec.
V we provide the conclusions.

\section{Quantum non-Markovianity}

Here we briefly review the main characteristics of the different approaches
to quantum non-Markovianity.

\subsection{Non-operational approach}

If the open system is not affected or perturbed during its evolution, the
unique object that allows to defining the presence or absence of memory
effects is its (unperturbed) density matrix propagator. The rigorous theory
of quantum dynamical semigroups \cite{alicki} motivate to associating the
(memoryless) quantum Markovian regime with propagators whose time-evolution
obey a Lindblad equation (or Gorini-Kossakowski-Sudarshan-Lindblad
equation). Consequently, any (scalar) measure or property that quantifies
departures of the system propagator from a Lindblad equation can be taken as
a witness of quantum memory effects.

Lindblad equations lead to completely positive propagators between two
arbitrary times \cite{alicki}. As is well known, completely positive
transformations lead to very specific contractive properties for different
distance measures and entropic quantities \cite{nielsen}. For example, the
TD between two arbitrary density matrixes $\rho $ and $\sigma ,$ defined as $%
\mathbf{D}(\rho ,\sigma )\equiv (1/2)\mathrm{Tr}|\rho -\sigma |,$ under a
completely positive transformation $\Phi $ fulfills the inequality $\mathbf{D%
}(\Phi \lbrack \rho ],\Phi \lbrack \sigma ])\leq \mathbf{D}(\rho ,\sigma ).$
Consequently, it is possible to \textit{define} quantum Markovianity by the
condition \cite{BreuerFirst}%
\begin{equation}
\mathbf{D}(\rho _{t+\tau }^{s},\sigma _{t+\tau }^{s})\leq \mathbf{D}(\rho
_{t}^{s},\sigma _{t}^{s}),  \label{Contraction}
\end{equation}%
where $\rho _{t}^{s}$ and $\sigma _{t}^{s}$ are two arbitrary evolved system
states that differ in their initial conditions, $\rho _{0}^{s}\neq \sigma
_{0}^{s}.$ Alternatively, one can interpret that quantum memory effects are
present whenever this inequality is not fulfilled for a set of two arbitrary
time intervals $t\geq 0$ and $\tau >0.$

In spite of the simplicity and efficacy of the previous theoretical frame,
in general it is not possible to know or infer which physical processes are
involved when the contractive condition~(\ref{Contraction}) is not
fulfilled. A remarkable advance in this direction was recently obtained in
Ref.~\cite{EntroBack} by establishing the inequality%
\begin{eqnarray}
\mathbf{D}(\rho _{t+\tau }^{s},\sigma _{t+\tau }^{s}) &\leq &\mathbf{D}(\rho
_{t}^{s},\sigma _{t}^{s})+\mathbf{D}(\rho _{t}^{e},\sigma _{t}^{e})
\label{General} \\
&&\!\!\!\!\!+\mathbf{D}(\rho _{t}^{se},\rho _{t}^{s}\otimes \rho _{t}^{e})+%
\mathbf{D}(\sigma _{t}^{se},\sigma _{t}^{s}\otimes \sigma _{t}^{e}).  \notag
\end{eqnarray}%
Here, $\rho _{t}^{se}$ and $\sigma _{t}^{se}$ are the evolved
system-environment states with initial conditions $\rho _{0}^{se}=\rho
_{0}^{s}\otimes \rho _{0}^{e}$ and $\sigma _{0}^{se}=\sigma _{0}^{s}\otimes
\sigma _{0}^{e}.$ As usual, the system and bath states follows from a
partial trace operation, $\rho _{t}^{s}=\mathrm{Tr}_{e}[\rho _{t}^{se}]$ and 
$\rho _{t}^{e}=\mathrm{Tr}_{s}[\rho _{t}^{se}].$ The asymmetry between
system and environment $(s\nleftrightarrow e)$ is introduced by taking in
both cases the same initial environmental state, $\rho _{0}^{e}=\sigma
_{0}^{e}.$

The result~(\ref{General}) only relies on the triangle inequality fulfilled
by the TD. Thus, it is valid for arbitrary system-environment models. In
addition, this expression allows to bounding the environment-to-system
backflow of information \textit{defined} by the \textquotedblleft
revivals\textquotedblright 
\begin{equation}
\mathbf{D}(\rho _{t+\tau }^{s},\sigma _{t+\tau }^{s})-\mathbf{D}(\rho
_{t}^{s},\sigma _{t}^{s})>0.  \label{Revivals}
\end{equation}%
The remaining (bounding) contributions in the rhs of Eq.~(\ref{General})
have a \textit{clear physical interpretation}. One can relate the
contribution $\mathbf{D}(\rho _{t}^{e},\sigma _{t}^{e})$ to changes in the
environmental state, while the terms $\mathbf{D}(\rho _{t}^{se},\rho
_{t}^{s}\otimes \rho _{t}^{e})+\mathbf{D}(\sigma _{t}^{se},\sigma
_{t}^{s}\otimes \sigma _{t}^{e})$ measure the correlations established
between the system and the environment~\cite{EntroBack}.
Nevertheless, it is important to realize that these physical processes do
not guarantee the developing of revivals. The right conclusion is that 
\textit{given} that there exists revivals, their origin can related to
changes in the environmental state or to the establishing of
system-environment correlations.

It was also proven that the inequality~(\ref{General}) remains valid when
the TD is replaced by a telescopic relative entropy and the square root of a
quantum Jensen-Shannon divergence~\cite{EntroBack}. Thus, the interpretation
of the bounds remains the same when using these entropic quantities.

\subsection{Operational approach}

In a probabilistic frame, given a sequence of system states $x\rightarrow
y\rightarrow z$ with joint probability $P(z,y,x),$ Markovianity is defined
by the condition%
\begin{equation}
P(z,y,x)=P(z|y)P(y|x)P(x),  \label{MarkovProb}
\end{equation}%
where $P(b|a)$ denotes in general the conditional probability of $b$ given $%
a.$ By Bayes rule, the equality~(\ref{MarkovProb}) implies the (memoryless)
condition $P(z|y,x)=P(z|y).$ Similar constraints emerge when considering
higher joint probabilities involving an arbitrary number of events~\cite%
{vanKampen}.

For quantum systems, the definition of Markovianity in terms of
probabilities unavoidably implies performing a set of system measurement
processes. In Ref.~\cite{modi}, by means of a process tensor formalism, the
Markovian condition is taken into account for arbitrary (higher order) joint
probabilities. Nevertheless, for \textit{quantum} systems coupled to
standard environment models (standard classical noises and/or unitary
system-environment interaction models), only three measurement events are
enough for detecting departures from a (probabilistic) Markovian regime~\cite%
{budiniCPF}. In such a case, the condition~(\ref{MarkovProb}) can be
conveniently rewritten as a CPF independence,%
\begin{equation}
P(z,x|y)=P(z|y)P(x|y).  \label{CPFIndep}
\end{equation}%
This result follows straightforwardly by using that $P(z,x|y)=P(z,y,x)/P(y),$
where $P(y)=\sum_{z,x}P(z,y,x).$

The CPF independence~(\ref{CPFIndep}) implies that any\ (conditional)
correlation between past and future events witnesses memory effects.
Correspondingly, a CPF correlation is defined as~\cite{budiniCPF}%
\begin{equation}
C_{pf}(t,\tau )|_{\breve{y}}\overset{d/r}{=}\sum_{z,x}zx[P(z,x|\breve{y}%
)-P(z|\breve{y})P(x|\breve{y})],  \label{CPFCorre}
\end{equation}%
where $\{x\}$ and $\{z\}$ are the (past and future) measurement outcomes.
The time dependence $(t,\tau )$ emerges because the past, present, and
future measurements are performed at the initial time $t=0,$ at time $t,$
and $t+\tau $ respectively. Evidently $C_{pf}(t,\tau )|_{\breve{y}}$
vanishes in a (probabilistic) Markovian regime [Eq.~(\ref{CPFIndep})].

In Eq.~(\ref{CPFCorre}) it was introduced the change $y\rightarrow \breve{y}%
, $ which is stretchy related with the definition of memory effects and
information flows in this approach. Two different measurement schemes are
necessary \cite{BIF}. In a deterministic scheme (denoted with the supra $d),$
after the intermediate measurement (whose outcome defines the conditional
property) none change is introduced. Hence, $\breve{y}=y.$ In a random
scheme (denoted with the supra $r),$ after the intermediate measurement the
system state is randomly chosen $(y\rightarrow \breve{y})$ over the set of
possible states associated to the outcomes $\{y\}.$ The CPF correlation is
defined with this renewed conditional state.

In the deterministic scheme, the CPF correlation $[C_{pf}(t,\tau )|_{\breve{y%
}}\overset{d}{\neq }0]$ detects memory effects [departures with respect to
Eq.~(\ref{MarkovProb}), or equivalently Eq.~(\ref{CPFIndep})] independently
of the specific system-environment model. In the random scheme, a
non-vanishing CPF correlation $[C_{pf}(t,\tau )|_{\breve{y}}\overset{r}{\neq 
}0],$ by \textit{definition,} detects the presence of environment-to-system
backflows of information (or bidirectional system-environment information
flows). This relation is motivated by the complementary case $C_{pf}(t,\tau
)|_{\breve{y}}\overset{r}{=}0$ that applies when the environment (which
induces the memory effects $C_{pf}(t,\tau )|_{\breve{y}}\overset{d}{\neq }0)$
is unperturbed by its coupling with the system~\cite{BIF}.

The previous characteristics of the deterministic and random schemes can be
easily understood from the properties of projective measurements performed
over bipartite systems \cite{casual}. Interestingly, the formalism remains
the same and is also valid for purely (classically) incoherent
system-environment arrangements.

\subsection{Bipartite propagator vs. single propagator}

Before comparing both approaches (next section), here we clarify which
dynamical objects determine each one. In the non-operational approach, the
presence of memory effects [TD revivals defined by Eq.~(\ref{Revivals})] can
be determined after knowing solely the system (single) propagator. In
contrast, for determining the bound defined by Eq.~(\ref{General}) it is
necessary to know the bipartite system-environment propagator specified for
a given initial bath state.

In contrast, the operational approach can only be characterized by knowing
(exact or approximate) the bipartite propagator  for different
initial bath states (the initial one and the bath state after the
intermediate measurement). As a matter of fact, the CPF correlation~(\ref%
{CPFCorre}) can be written as a function of the joint probability $P(z,%
\breve{y},x).$ Assuming that the three measurements are projective ones, in
the deterministic scheme it reads~\cite{BIF}%
\begin{equation}
\frac{P(z,\breve{y},x)}{P(x)}\overset{d}{=}\mathrm{Tr}_{se}(E_{z}\mathcal{G}%
_{t+\tau ,t}^{se}[\rho _{\breve{y}}\otimes \mathrm{Tr}_{s}(E_{\breve{y}}%
\mathcal{G}_{t,0}^{se}[\rho _{x}^{se}])]),  \label{Joint_d}
\end{equation}%
while in the random scheme it is~\cite{BIF}%
\begin{equation}
\frac{P(z,\breve{y},x)}{P(x)}\overset{r}{=}\mathrm{Tr}_{se}(E_{z}\mathcal{G}%
_{t+\tau ,t}^{se}[\rho _{\breve{y}}\otimes \mathrm{Tr}_{s}(\mathcal{G}%
_{t,0}^{se}[\rho _{x}^{se}])])\wp (\breve{y}|x).  \label{Joint_r}
\end{equation}%
In these expressions, $\mathcal{G}_{t+\tau ,t}^{se}$ is the \textit{%
bipartite propagator} between $t$ and $t+\tau .$ In addition, $E_{m}\equiv
|m\rangle \langle m|$ and $\rho _{m}\equiv |m\rangle \langle m|$ $[m=z,%
\breve{y},x]$ represent the (positive) effect measurement operators and
post-measurement states respectively. The set $\{|m\rangle \}$ $[m=z,\breve{y%
},x]$ are the eigenstates of each measured observable. Furthermore, $\rho
_{x}^{se}\equiv \rho _{x}\otimes \rho _{0}^{e}$ and $P(x)=\langle x|\rho
_{0}^{s}|x\rangle .$ The random scheme is parametrized by an arbitrary
conditional probability $\wp (\breve{y}|x)$ that defines the change in the
system state $(y\rightarrow \breve{y})$ after the intermediate measurement.

The different dependence of both approaches on the bipartite propagator
leads to strong different conclusions about memory effects and information
flows, which are analyzed in the next section.

\section{Comparing both approaches}

In order to perform a systematic comparison we consider different
system-environment models and approximations. In general, we assume that the
bipartite system-environment state $\rho _{t}^{se}$ evolves as%
\begin{equation}
\frac{d}{dt}\rho _{t}^{se}=(\mathcal{L}_{s}+\mathcal{L}_{e}+\mathcal{L}%
_{se})[\rho _{t}^{se}],  \label{BipartiteEvolution}
\end{equation}%
where $\mathcal{L}_{s}$ and $\mathcal{L}_{e}$ define the self-dynamics of
the system and the environment respectively, while $\mathcal{L}_{se}$\
defines their mutual interaction. This interaction term may be unitary or
includes dissipative couplings.

\subsection{Born-Markov approximation}

For systems weakly coupled to their environments, the Born-Markov
approximation \cite{breuerbook} allows to write the bipartite state as%
\begin{equation}
\rho _{t}^{se}\simeq \rho _{t}^{s}\otimes \rho _{0}^{e},  \label{BMA}
\end{equation}%
where $\rho _{t}^{s}$ is the system state, while $\rho _{0}^{e}$ is the
(almost) unperturbed environment state.

When this approximation is valid, in the \textit{non-operational approach},
it is simple to check that Eq.~(\ref{General}) reduces to Eq.~(\ref%
{Contraction}). In fact, $\mathbf{D}(\rho _{t}^{e},\sigma _{t}^{e})=\mathbf{D%
}(\rho _{t}^{se},\rho _{t}^{s}\otimes \rho _{t}^{e})=\mathbf{D}(\sigma
_{t}^{se},\sigma _{t}^{s}\otimes \sigma _{t}^{e})=0.$ Furthermore, $\rho
_{t}^{s}$ can be well approximated by a Lindblad equation, which guarantees
the absence of any revival in $\mathbf{D}(\rho _{t}^{s},\sigma _{t}^{s}).$
Thus, the dynamics is Markovian.

In the \textit{operational approach}, by introducing the approximation~(\ref%
{BMA}) into Eqs.~(\ref{Joint_d}) and (\ref{Joint_r}) straightforwardly it
follows that $C_{pf}(t,\tau )|_{\breve{y}}\overset{d/r}{=}0$ [Eq.~(\ref%
{CPFCorre})]. These results are independent of which observables are
measured. Thus, the dynamics is Markovian.

In this case [Eq.~(\ref{BMA})] both approaches coincides. Strong differences
appear in the cases studied below.

\subsection{Casual bystander environments}

A wide class of \textquotedblleft non-Markovian\textquotedblright\ dynamics
can be derived by assuming that the system interacts with a
\textquotedblleft casual bystander\textquotedblright\ environment. These
baths are defined by the independence of their marginal states $\rho
_{t}^{e}=\mathrm{Tr}_{s}[\rho _{t}^{se}]$ of any degree of freedom of the
system. Alternatively, the time evolution of $\rho _{t}^{e}$ can be written
in the environment Hilbert space without involving any operator or state of
the system. These properties must be valid for arbitrary system and
environment (separable) initial conditions.

For fulfilling the previous properties, the interaction term $\mathcal{L}%
_{se}$ in the general evolution~(\ref{BipartiteEvolution}) must be
restricted such that%
\begin{equation}
\mathrm{Tr}_{s}(\mathcal{L}_{se}[\rho _{t}^{se}])=\mathcal{A}[\rho _{t}^{e}],
\label{CBE}
\end{equation}%
where $\mathcal{A}$\ is an arbitrary superoperator acting on $\rho _{t}^{e}$
that does not have any dependence on the system degrees of freedom. In
general, this constraint can only be satisfied by dissipative (non-unitary)
system-environment couplings. On the other hand, the bath dynamics can be
quantum \cite{casual} or a classical (incoherent) one~\cite{maximal}.

In the \textit{non-operational approach}, the independence of the
environment state on the system degrees of freedom cannot be translated to
any restriction on the inequality defined by Eq.~(\ref{General}). In fact,
under the constraint~(\ref{CBE}), the TD may or not present revivals, 
\textit{property that can only be cheeked for each specific model}. Thus,
some dynamics are classified as Markovian and other as non-Markovian. The
unique simplification that can be introduced is to assume that the
environment state does not evolve in time, $\rho _{t}^{e}=\rho _{0}^{e},$
that is, the environment begins in its stationary state. In this case, Eq.~(%
\ref{General}) reduces to%
\begin{eqnarray}
\mathbf{D}(\rho _{t+\tau }^{s},\sigma _{t+\tau }^{s})-\mathbf{D}(\rho
_{t}^{s},\sigma _{t}^{s}) &\leq &\mathbf{D}(\rho _{t}^{se},\rho
_{t}^{s}\otimes \rho _{0}^{e})  \notag \\
&&\!\!\!\!\!+\mathbf{D}(\sigma _{t}^{se},\sigma _{t}^{s}\otimes \rho
_{0}^{e}).  \label{TdStacion}
\end{eqnarray}%
Even in this case $(\rho _{t}^{e}=\rho _{0}^{e}),$ the TD may or not present
revivals, that is, depending on the model, the system may be classified as
Markovian or non-Markovian.

In Eq. (\ref{TdStacion}), any environment-to-system backflow of information
can be related to the establishing of the correlations $\mathbf{D}(\rho
_{t}^{se},\rho _{t}^{s}\otimes \rho _{0}^{e})+\mathbf{D}(\sigma
_{t}^{se},\sigma _{t}^{s}\otimes \rho _{0}^{e}).$ Certainly, the
system-environment correlations (always) changes in time. 
Nevertheless, even when there is not revivals in the TD system-environment
correlations are established. This feature represent a central problem for
the interpretation of this approach. In addition, here the environment
state is completely independent of the system (and even of time). Thus, the
revivals of the TD must be taken as a (mathematical) model-dependent
property whose origin cannot be related to any physical process that implies
a \textit{physical} transfer of information from the environment to the
system.

A different perspective emerges in the \textit{operational approach}. By
using the independence of the environment state $[\rho _{t}^{e}=\mathrm{Tr}%
_{s}(\rho _{t}^{se})]$ of any degree of freedom of the system, it is
possible to check that the joint probability~(\ref{Joint_d}) of the
deterministic scheme \textit{does not fulfill} the Markov property~(\ref%
{MarkovProb}). In contrast, it is simple to check that the joint
probability~(\ref{Joint_r}) of the random scheme \textit{fulfill} the Markov
property~(\ref{MarkovProb}). Consequently, a casual bystander environment
leads to\ the CPF correlations [Eq.~(\ref{CPFCorre})]%
\begin{equation}
C_{pf}(t,\tau )|_{\breve{y}}\overset{d}{\neq }0,\ \ \ \ \ \ \ \ \ \
C_{pf}(t,\tau )|_{\breve{y}}\overset{r}{=}0.  \label{CPFforCBE}
\end{equation}%
In this approach, the property $C_{pf}(t,\tau )|_{\breve{y}}\overset{d}{\neq 
}0,$ valid for any model under the constraint~(\ref{CBE}), implies that the
system dynamics is non-Markovian.  Its origin can be related to
the establishing of (arbitrary) system-environment correlations. On the
other hand, the property $C_{pf}(t,\tau )|_{\breve{y}}\overset{r}{=}0,$
which is valid for arbitrary measurement processes and specific models, is
read as the \textit{absence} of bidirectional system-environment information
flows. In fact, given that the environment is characterized by a
self-dynamics that is completely independent of the system, any
environment-to-system backflow of information (as detected in the
non-operational approach) does not rely on any physical process that affect
the environment state neither its dynamics.

The meaning of the previous analysis is clarified by specifying different
bipartite models that fulfill the evolution (\ref{BipartiteEvolution}) and
the constraint~(\ref{CBE}).

\subsubsection{Classical mixture of quantum Markovian dynamics}

Given a set of \textit{different} system Lindblad superoperators $\{\mathcal{%
L}_{s}^{c}\},$ which may include both unitary and dissipative contributions,
and given a set of normalized positive weights $\{p_{c}\},$ $%
\sum_{c}p_{c}=1, $ a classical statistical mixture of Markovian dynamics is
defined by the bipartite state%
\begin{equation}
\rho _{t}^{se}=\sum_{c}\exp (t\mathcal{L}_{s}^{c})[\rho _{0}^{s}]\otimes
p_{c}|c\rangle \langle c|.  \label{Superposition}
\end{equation}%
Here, $\{|c\rangle \langle c|\}$ is a set of projectors associated to the
environment space. The marginal system and environment states read%
\begin{equation}
\rho _{t}^{s}=\sum_{c}p_{c}\exp (t\mathcal{L}_{s}^{c})[\rho _{0}^{s}],\ \ \
\ \ \ \rho _{t}^{e}=\sum_{c}p_{c}|c\rangle \langle c|.  \label{SystemMixture}
\end{equation}%
Memory effects in this kind of non-Markovian system dynamics has been
explored in the literature \cite{poland}. Notice that the environment does
not have any dynamics. Even more, the system dynamics can be performed by
mixing in a random way (with weight $p_{c})$ each of the evolved Markovian
system states $\exp (t\mathcal{L}_{s}^{c})[\rho _{0}^{s}].$ Thus, the
detection of an environment-to-system backflow of information via Eq.~(\ref%
{Revivals}) seems to have a formal mathematical interpretation rather than a
physical one. On the other hand, in the operational approach this case is
characterized by Eq.~(\ref{CPFforCBE}), which guaranty the presence of
memory effects $C_{pf}(t,\tau )|_{\breve{y}}\overset{d}{\neq }0,$ but not
any bidirectional information flow, $C_{pf}(t,\tau )|_{\breve{y}}\overset{r}{%
=}0.$

\subsubsection{Interaction with stochastic classical degrees of freedom}

When the system interacts with stochastic classical degrees of freedom, the
bipartite state can be written as%
\begin{equation}
\rho _{t}^{se}=\sum_{c}\rho _{t}^{c}\otimes p_{c}(t)|c\rangle \langle c|.
\label{BipartitoClassical}
\end{equation}%
In contrast to the previous case [Eq.~(\ref{Superposition})], the weights $%
\{p_{c}(t)\}$ are time-dependent and the evolution of the states $\{\rho
_{t}^{c}\}$ may involve coupling between all of them. In fact, under the
constraint~(\ref{CBE}), the more general evolution can be written as~\cite%
{maximal}%
\begin{equation}
\frac{d\tilde{\rho}_{t}^{c}}{dt}=\mathcal{L}_{s}^{c}[\tilde{\rho}%
_{t}^{c}]-\sum_{c^{\prime }}\gamma _{c^{\prime }c}\tilde{\rho}%
_{t}^{c}+\sum_{c^{\prime }}\gamma _{cc^{\prime }}\mathbb{S}_{cc^{\prime }}[%
\tilde{\rho}_{t}^{c^{\prime }}].  \label{EvolveClassical}
\end{equation}%
Here, $\tilde{\rho}_{t}^{c}\equiv p_{c}(t)\rho _{t}^{c}.$ Thus, $p_{c}(t)=%
\mathrm{Tr}_{s}(\tilde{\rho}_{t}^{c}).$ Furthermore, $\{\mathbb{S}%
_{cc^{\prime }}\}$ are arbitrary completely positive system transformations,
which are trace preserving, $\mathrm{Tr}_{s}(\mathbb{S}_{cc^{\prime }}[\rho
])=\mathrm{Tr}_{s}(\rho ).$ Consequently, the environment probabilities $%
\{p_{c}(t)\}$ obey a classical master equation%
\begin{equation}
\frac{dp_{c}(t)}{dt}=-\sum_{c^{\prime }}\gamma _{c^{\prime
}c}p_{c}(t)+\sum_{c^{\prime }}\gamma _{cc^{\prime }}p_{c^{\prime }}(t),
\end{equation}%
which in turn shows the role played by the coupling rates $\{\gamma
_{c^{\prime }c}\}.$ In contrast, the system dynamics depart from a Markovian
(Lindblad) evolution. From some specific models, it is possible to recover
some phenomenological non-Markovian master equations (see for example \cite%
{lidar}).

In the non-operational\ approach it is very difficult to predict if a given
dynamics [Eq.~(\ref{BipartitoClassical})] lead or not to revivals in the TD.
If the incoherent degrees of freedom begin in their stationary state, $%
p_{c}(0)=\lim_{t\rightarrow \infty }p_{c}(t),$ one is confronted with the
bounds defined by Eq.~(\ref{TdStacion}). Even in this case, one cannot
predict when there exist or not an environment-to-system backflow of
information.

Interestingly, the origin of the contributions $\mathbf{D}(\rho
_{t}^{se},\rho _{t}^{s}\otimes \rho _{0}^{e})+\mathbf{D}(\sigma
_{t}^{se},\sigma _{t}^{s}\otimes \rho _{0}^{e})$ in Eq.~(\ref{TdStacion})
[or in general in Eq.~(\ref{General})] can be easily read from the
evolution~(\ref{EvolveClassical}). In fact, this equation shows that the
system evolution is \textit{totally conditioned} to the environment
dynamics. The contributions $\mathcal{L}_{s}^{c}$ are\ \textquotedblleft
active\textquotedblright\ whenever the environment is in the state $%
|c\rangle \langle c|.$ Furthermore, the system suffers the transformation $%
\rho \rightarrow \mathbb{S}_{c^{\prime }c}[\rho ]$ whenever the environment
\textquotedblleft jumps\textquotedblright\ between the states $c\rightarrow
c^{\prime }.$ This is the physical mechanism that leads to the
system-environment correlations, which in turn does not imply any
system-dependent change in the environment state or dynamics. Thus, the
interpretation of revivals in the TD as environment-to-system backflow of
information is again controversial.

Independently of the Lindblad contributions $\{\mathcal{L}_{s}^{c}\},$ the
superoperators $\{\mathbb{S}_{cc^{\prime }}\},$ and rates $\{\gamma
_{c^{\prime }c}\},$ the operational approach is characterized by Eq.~(\ref%
{CPFforCBE}), that is, the dynamics is non-Markovian $[C_{pf}(t,\tau )|_{%
\breve{y}}\overset{d}{\neq }0]$ without the developing of any bidirectional
system-environment information flow $[C_{pf}(t,\tau )|_{\breve{y}}\overset{r}%
{=}0].$

\subsubsection{Environmental quantum degrees of freedom}

The condition Eq.~(\ref{CBE}) can be satisfied even when the environment is
a quantum one, that is, it develops coherent behaviors. In this case, the
bipartite state can be written as%
\begin{equation}
\rho _{t}^{se}=\sum \rho _{t}^{c}\otimes p_{c}(t)|c_{t}\rangle \langle
c_{t}|.  \label{BipartitoCBE}
\end{equation}%
In contrast to Eq.~(\ref{BipartitoClassical}), due to the quantum nature of
the environment, the projectors $\{|c_{t}\rangle \langle c_{t}|\}$ are
time-dependent. In fact, they define the base in which the environment
density matrix $\rho _{t}^{e}$ is diagonal. The more general bipartite
evolution~(\ref{BipartiteEvolution}) under the constraint~(\ref{CBE}), in
its diagonal representation, is given by \cite{casual}%
\begin{eqnarray}
\frac{d}{dt}\rho _{t}^{se} &=&(\mathcal{L}_{s}+\mathcal{L}_{e})[\rho
_{t}^{se}]+\sum_{\alpha }\Gamma _{\alpha }\ B_{\alpha }\mathbb{S}_{\alpha
}[\rho _{t}^{se}]B_{\alpha }^{\dag }  \notag \\
&&-\frac{1}{2}\sum_{\alpha }\Gamma _{\alpha }\{B_{\alpha }^{\dag }B_{\alpha
},\rho _{t}^{se}\}_{+},  \label{GeneralSU}
\end{eqnarray}%
where $\{\cdot ,\cdot \}_{+}$ is an anticommutator operation. Furthermore, $%
\{B_{\alpha }\}$ are arbitrary environment operators, while $\mathbb{S}%
_{\alpha }$ are completely positive trace-preserving system superoperators.
The rates $\{\Gamma _{\alpha }\}$ set the environment dynamics. In fact,%
\begin{equation}
\frac{d}{dt}\rho _{t}^{e}=\mathcal{L}_{e}[\rho _{t}^{e}]+\sum_{\alpha
}\Gamma _{\alpha }\ (B_{\alpha }\rho _{t}^{e}B_{\alpha }^{\dag }-\frac{1}{2}%
\{B_{\alpha }^{\dag }B_{\alpha },\rho _{t}^{e}\}_{+}),  \label{BathLindblad}
\end{equation}%
which is a Lindblad dynamics completely independent of the system degrees of
freedom. These evolutions recover, as particular cases, some
phenomenological collisional models introduced in the literature (see for
example \cite{collision}).

The physical interpretation of the evolution~(\ref{GeneralSU}) is quite
similar to that of Eq.~(\ref{EvolveClassical}). In fact, here the
application of the system superoperators $\mathbb{S}_{\alpha }$ occur
whenever the environment suffer a transition associated to the operators $%
B_{\alpha }.$ This (unidirectional) mechanism defines how the
system-environment correlations are build up.

In the non-operational approach, even when the environment begins in its
stationary state $\rho _{0}^{e}=\lim_{t\rightarrow \infty }\rho _{t}^{e}$
[where $\rho _{t}^{e}$ obeys Eq.~(\ref{BathLindblad})], it is not possible
to infer for an arbitrary model the presence or absence of revivals in the
TD [Eq.~(\ref{Revivals})]. In contrast, the operational approach is still
characterized by Eq.~(\ref{CPFforCBE}).

\subsection{Unitary system-environment interactions}

Independently of the specific models, the correlation between the system and
the casual bystander environments introduced previously does
not involve quantum entanglement \cite{entanglement} [see the separable
states Eqs.~(\ref{Superposition}), (\ref{BipartitoClassical}), and (\ref%
{BipartitoCBE})]. In contrast, quantum entanglement may emerges when
considering Hamiltonian (time-reversible) system-environment interactions. 
In fact, solely for special system-environment initial conditions a
bipartite unitary dynamics does not induce quantum entanglement \cite{katarzyna}.

The total Hamiltonian is written as%
\begin{equation}
H_{T}=H_{s}+H_{e}+H_{I}.  \label{Htotal}
\end{equation}%
Each contribution correspond respectively to the system, environment, and
interaction Hamiltonians respectively. The bipartite propagator is%
\begin{equation}
\mathcal{G}_{t,t_{0}}^{se}[\bullet ]=\exp [-i(t-t_{0})H_{T}]\bullet \exp
[+i(t-t_{0})H_{T}].  \label{Gunitario}
\end{equation}

In the\textit{\ non-operational approach}, each contribution in the rhs of
Eq.~(\ref{General}) make complete sense in this context. In fact, almost all
unitary interactions lead to a change in the environment state and also
induce the developing of (arbitrary) system-environment correlations. When
revivals in the TD develops, Eq.~(\ref{General}) defines a bound with a
clear physical meaning. Nevertheless, in general it is not possible to infer
which kind of dynamics develop or not revivals in the TD. Even for a given
(Hamiltonian) model, depending on the underlying parameters the system
dynamics may be Markovian or not. Consequently, it is not clear which
physical property defines the boundary between Markovian and non-Markovian
dynamics.

In the \textit{operational approach}, given that the state and dynamics of
the environment are in general modified by a unitary interaction, instead of
Eq.~(\ref{CPFforCBE}), here it follows%
\begin{equation}
C_{pf}(t,\tau )|_{\breve{y}}\overset{d}{\neq }0,\ \ \ \ \ \ \ \ \ \
C_{pf}(t,\tau )|_{\breve{y}}\overset{r}{\neq }0.  \label{CPFUnitary}
\end{equation}%
Both inequalities can be supported by performing a perturbation theory based
on projectors techniques \cite{bonifacio}. Consistently, it has been shown
that even close to the validity of a Born-Markov approximation the
operational approach can detect memory effects \cite{rio}.

The inequality $C_{pf}(t,\tau )|_{\breve{y}}\overset{d}{\neq }0$ implies
that the system dynamics is non-Markovian (system-environment correlations
are developed during the evolution), while $C_{pf}(t,\tau )|_{\breve{y}}%
\overset{r}{\neq }0$ detects the presence of bidirectional information
flows. In fact, here the environment state and evolution always depend on
the system degrees of freedom.

There exist a unique exception to Eq.~(\ref{CPFUnitary}), which reduces to
Eq.~(\ref{CPFforCBE}). Hence, even when the environment state is modified,
for any system observables one obtain $C_{pf}(t,\tau )|_{\breve{y}}\overset{r%
}{=}0.$ While this property is certainly\textit{\ undesirable}, this case
has a clear physical interpretation. It emerges when, in a given
environmental base $\{|e\rangle \},$ the diagonal part of the bipartite
propagator~(\ref{Gunitario}) can be written as%
\begin{equation}
\langle e|\mathcal{G}_{t,0}^{se}[\bullet ]|e\rangle =\mathcal{T}%
_{t,0}^{(e)}\langle e|\bullet |e\rangle ,  \label{DiagonalCondition}
\end{equation}%
where $\mathcal{T}_{t,0}^{(e)}$ is a \textit{system} (density matrix)
propagator that parametrically depends on each environmental state $%
\{|e\rangle \}.$ The condition~(\ref{DiagonalCondition}) is fulfilled, 
\textit{for example}, when the environment and interaction Hamiltonians
commutate%
\begin{equation}
\lbrack H_{e},H_{I}]=0.  \label{Conmutador}
\end{equation}

Introducing the condition~(\ref{DiagonalCondition}) into Eqs.~(\ref{Joint_d}%
) and~(\ref{Joint_r}) it is possible to check that $C_{pf}(t,\tau )|_{\breve{%
y}}\overset{d}{\neq }0,$\ and $C_{pf}(t,\tau )|_{\breve{y}}\overset{r}{=}0.$
This last equality \textit{does not imply} that the environment in not
affected. It emerges because the system state assumes the structure%
\begin{equation}
\rho _{t}^{s}=\mathrm{Tr}_{e}(\mathcal{G}_{t,0}^{se}[\rho _{0}^{s}\otimes
\rho _{0}^{e}])=\sum_{e}\langle e|\rho _{0}^{e}|e\rangle \mathcal{T}%
_{t,0}^{(e)}[\rho _{0}^{s}].  \label{SystemRandomUnitary}
\end{equation}%
Therefore, the system evolution can be written as a statistical
superposition of unitary maps, quite similar to Eq.~(\ref{SystemMixture}).
In consequence, for unitary system-environment models the condition $%
C_{pf}(t,\tau )|_{\breve{y}}\overset{r}{=}0$ allows to detect when the
system dynamics (even between measurements) can be represented by a \textit{%
Hamiltonian ensemble}, property that has been of interest in the recent
literature~\cite{nori}.

\section{Example}

In this section we consider an explicit example of the dynamics discussed
previously. The quantum system $(s),$ taken for simplicity as a two-level
system, interacts with an incoherent environment $(e)$ (see Sect. 3.2.2),
which here is defined by four discrete states denoted as $\{|1\rangle
,|2\rangle ,|3\rangle ,|4\rangle \}.$ Correspondingly, the bipartite
system-environment state is written as%
\begin{equation}
\rho _{t}^{se}=\sum_{k=1,2,3,4}\tilde{\rho}_{k}(t)\otimes |k\rangle \langle
k|.  \label{Bipartita}
\end{equation}%
The system and environment states then read%
\begin{equation}
\rho _{t}^{s}=\sum_{k=1,2,3,4}\tilde{\rho}_{k}(t),\ \ \ \ \ \ \ \rho
_{t}^{e}=\sum_{k=1,2,3,4}p_{k}(t)|k\rangle \langle k|,  \label{RhoSystema}
\end{equation}%
where $p_{k}(t)=\mathrm{Tr}_{s}[\tilde{\rho}_{k}(t)].$ The evolution of the
unnormalized system states $\{\tilde{\rho}_{k}(t)\}_{k=1}^{k=4}$ is taken as 
\begin{subequations}
\label{EvolucionRhoTilde}
\begin{eqnarray}
\frac{d\tilde{\rho}_{4}(t)}{dt} &=&-\gamma \tilde{\rho}_{4}(t)+\phi
\sum_{k=1,2,3}\sigma _{k}\tilde{\rho}_{k}(t)\sigma _{k}, \\
\frac{d\tilde{\rho}_{k}(t)}{dt} &=&-\phi \tilde{\rho}_{k}(t)+\left( \frac{%
\gamma }{3}\right) \sigma _{k}\tilde{\rho}_{4}(t)\sigma _{k},\ \ \ k=1,2,3.\
\ \ \ \ \ \ \ 
\end{eqnarray}%
In this expression, $\gamma $ and $\phi $\ are characteristic coupling
rates. Furthermore, the set of Pauli matrixes is denoted as $(\sigma
_{x},\sigma _{y},\sigma _{z},\mathrm{I})\leftrightarrow (\sigma _{1},\sigma
_{2},\sigma _{3},\sigma _{4}),$ where\ $\mathrm{I}$ is the identity\ matrix
in the two-dimensional system Hilbert space. From Eq.~(\ref%
{EvolucionRhoTilde}), the evolution of the environment populations is
defined by the following classical master equation 
\end{subequations}
\begin{subequations}
\label{Poblaciones}
\begin{eqnarray}
\frac{dp_{4}(t)}{dt} &=&-\gamma p_{4}(t)+\phi \sum_{k=1,2,3}p_{k}(t), \\
\frac{dp_{k}(t)}{dt} &=&-\phi p_{k}(t)+\left( \frac{\gamma }{3}\right)
p_{4}(t),\ \ \ k=1,2,3.\ 
\end{eqnarray}%
This equation is completely independent of the system degrees of freedom.
Thus, the evolution (\ref{EvolucionRhoTilde}) has a simple interpretation.
When the environment suffer the transition $|4\rangle \overset{\gamma /3}{%
\rightarrow }|k\rangle $ or the transition $|k\rangle \overset{\phi }{%
\rightarrow }|4\rangle $ $(k=1,2,3),$ the transformation $\sigma _{k}\bullet
\sigma _{k}$ is \textit{conditionally} applied over the open quantum system.

Eq.~(\ref{EvolucionRhoTilde}) can be solved after specifying the bipartite
initial conditions. We consider a separable state, $\rho _{0}^{se}=\rho
_{0}^{s}\otimes \rho _{0}^{e},$ which implies $\tilde{\rho}_{k}(0)=\rho
_{0}^{s}p_{k}(0).$ In general, each auxiliary state $\tilde{\rho}_{k}(t)$
can be written as a superposition of the Pauli channels acting on the
initial system state $\rho _{0}^{s},$ that is, 
\end{subequations}
\begin{equation}
\tilde{\rho}_{k}(t)=\sum_{j=1,2,3,4}g_{k}^{j}(t)\sigma _{j}\rho
_{0}^{s}\sigma _{j},  \label{RhoTildeK}
\end{equation}%
where $\{g_{k}^{j}(t)\}$ are (sixteen) scalar functions that depend on time.
Their initial conditions are $g_{k}^{4}(0)=p_{k}(0)$ and $g_{k}^{j}(0)=0,$
with $j=1,2,3,$ and $k=1,2,3,4.$ The evolution of the set $\{g_{k}^{j}(t)\}$
follows after inserting the previous expression for $\tilde{\rho}_{k}(t)$
into Eq.~(\ref{EvolucionRhoTilde}). Consistently with their definition, $%
p_{k}(t)=\mathrm{Tr}_{s}[\tilde{\rho}_{k}(t)],$ the environment populations
are recovered as%
\begin{equation}
p_{k}(t)=\sum_{j=1,2,3,4}g_{k}^{j}(t).
\end{equation}

\subsection{Depolarizing dynamics}

The evolution of the auxiliary states Eq.~(\ref{EvolucionRhoTilde}) is
(structurally) the same for the states $\{\tilde{\rho}_{1}(t),\tilde{\rho}%
_{2}(t),\tilde{\rho}_{3}(t)\}.$ Thus, if we consider environment initial
conditions where $p_{1}(0)=p_{2}(0)=p_{3}(0),$ from Eqs.~(\ref{RhoSystema})
and (\ref{RhoTildeK}) it follows that the solution map $\rho
_{0}^{s}\rightarrow \rho _{t}^{s}$ must be a depolarizing channel \cite%
{nielsen}, that is,%
\begin{equation}
\rho _{t}^{s}=w(t)\rho _{0}^{s}+\frac{1-w(t)}{3}\sum_{k=1,2,3}\sigma
_{k}\rho _{0}^{s}\sigma _{k},  \label{DepoRho}
\end{equation}%
where the positive weight $w(t),$\ from Eq.~(\ref{RhoTildeK}), follows as%
\begin{equation}
w(t)=\sum_{k=1,2,3,4}g_{k}^{4}(t).
\end{equation}%
Consistently, $[1-w(t)]/3=\sum_{k=1,2,3,4}g_{k}^{j}(t),$ with $j=1,$ $2,$ $%
3. $

The more natural initial conditions for the environment are their stationary
populations $p_{k}^{\infty }\equiv \lim_{t\rightarrow \infty }p_{k}(t),$
where $p_{k}(t)$ are defined by Eq.~(\ref{Poblaciones}). Straightforwardly,
we get 
\begin{equation}
p_{4}^{\infty }=\frac{\phi }{\gamma +\phi },\ \ \ \ \ \ \ \ p_{k}^{\infty }=%
\frac{1}{3}\frac{\gamma }{\gamma +\phi }\ \ \ \ \ (k=1,2,3).
\label{Estacionarias}
\end{equation}%
Under the assumption $p_{k}(0)=p_{k}^{\infty },$ after getting the set $%
\{g_{k}^{j}(t)\}$ in an explicit analytical way, the function $w(t)$ that
characterizes the depolarizing channel Eq.~(\ref{DepoRho}) can be written as%
\begin{equation}
w(t)=\frac{(\gamma ^{2}+3\phi ^{2})}{3(\gamma +\phi )^{2}}+\frac{4\gamma
\phi }{3(\gamma +\phi )^{2}}e^{-(\gamma +\phi )t}+\frac{2\gamma }{3(\gamma
+\phi )}e^{-\phi t},  \label{W(t)}
\end{equation}%
which consistently satisfies $w(0)=1.$ Furthermore, $\lim {}_{t\rightarrow
\infty }w(t)\neq 0.$ On the hand, the environment dynamics is stationary,
that is, $p_{k}(t)=p_{k}(0)=p_{k}^{\infty }$ [Eq.~(\ref{Estacionarias})].

\subsection{Operational vs. non-operational quantum non-Markovianity}

In the\textit{\ non-operational approach}, quantum non-Markovianity is
defined by the revivals in the trace distance between two different initial
states, Eq.~(\ref{Revivals}). By using that $(\mathrm{I}/2)=(\rho
+\sum_{k=1,2,3}\sigma _{k}\rho \sigma _{k})/4$ \cite{nielsen},\ the
depolarizing map (\ref{DepoRho}) can be rewritten as $\rho _{t}^{s}=w(t)\rho
_{0}^{s}+(1/3)[1-w(t)](2\mathrm{I}-\rho _{0}^{s}).$ Thus, the trace distance
straightforwardly can be written as%
\begin{equation}
D[\rho _{t}^{s},\sigma _{t}^{s}]=\left\vert \frac{4w(t)-1}{3}\right\vert
D[\rho _{0}^{s},\sigma _{0}^{s}]\equiv d(t)D[\rho _{0}^{s},\sigma _{0}^{s}]
\label{TraceDistance}
\end{equation}%
where $D[\rho _{0}^{s},\sigma _{0}^{s}]$ is the trace distance between the
two initial states $\rho _{0}^{s}$ and $\sigma _{0}^{s}.$ Notice that the
decay of the trace distance does not depends on the initial states, being
dictated by the function $d(t).$

In Fig.~1(a) we plot the function $d(t)$ for different values of the
characteristic parameter $\phi /\gamma .$ As expected from Eq.~(\ref{W(t)}), 
$D[\rho _{t}^{s},\sigma _{t}^{s}]$ decay in a monotonous way without
developing any revival. Thus, under the trace distance criteria, the
dynamics is \textit{Markovian} and there is not any environment-to-system
backflow of information. Nevertheless, notice that for any value of $\phi
/\gamma $ system-environment correlations are built up during the dynamics
[see Eq.~(\ref{Bipartita})]. This feature, which is irrelevant for the TD
decay behavior, is relevant for the CPF correlation. 
\begin{figure}[tbp]
\includegraphics[bb=22 350 520 555,angle=0,width=8.5cm]{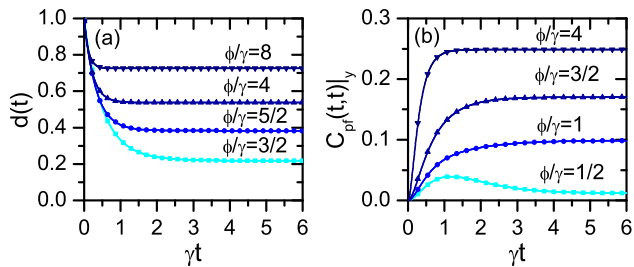}
\caption{(a) Decay of the trace distance $d(t)$ [Eq.~(\protect\ref%
{TraceDistance})] corresponding to the model (\protect\ref{EvolucionRhoTilde}%
). (b) Time dependence of the CPF correlation $C_{pf}(t,t)|_{\breve{y}}$ in
the deterministic scheme \protect\cite{CPF} corresponding to the same model.
The value of the quotient $\protect\phi /\protect\gamma $ is indicated in
each plot.}
\end{figure}

In the \textit{operational approach}, the presence of memory effects is
witnessed by the CPF correlation [Eq.~(\ref{CPFCorre})] in the deterministic
scheme. We assume that the three measurements are projective ones, all of
them being performed in the $z$-direction of the Bloch sphere. Furthermore
the initial condition of the system is taken as $\rho _{0}^{s}=|\psi \rangle
\langle \psi |,$ where $|\psi \rangle $ is an eigenstate of the $x$-Pauli
matrix. Explicit general expression for $C_{pf}(t,\tau )|_{\breve{y}}$ in
terms of the coefficients $\{g_{k}^{j}(t)\}$ can be found in Ref.~\cite%
{multi} (see corresponding Appendix D). Under the previous assumptions, the
CPF correlation can be obtained in an analytical way, which is written in 
\cite{CPF}. Simple expressions are obtained for specific values of the decay
rates. For example, for $\phi =\gamma $ it follows%
\begin{eqnarray}
&&C_{pf}(t,\tau )|_{\breve{y}}\overset{d}{=}\frac{4}{81}(1-e^{-\gamma
t})(1-e^{-\gamma \tau }) \\
&&\ \ \ \ \ \ \ \ \ \ \ \ \ \ \ \ \ \ \times (2+e^{-\gamma t}+e^{-\gamma
\tau }+5e^{-\gamma (t+\tau )}).  \notag
\end{eqnarray}%
Due to the symmetry of the problem, in all cases $C_{pf}(t,\tau )|_{\breve{y}%
}$ does not depend on the value of the conditional $\breve{y}=\pm 1.$ In
Fig.~1(b) we plot the CPF correlation at equal times, $C_{pf}(t,t)|_{\breve{y%
}},$ for different values of $\phi /\gamma .$ In contrast to the
non-operational approach, here for all possible values of the characteristic
parameter $\phi /\gamma $ it is fulfilled $C_{pf}(t,\tau )|_{\breve{y}}%
\overset{d}{\neq }0,$ which indicates a \textit{non-Markovian} regime. In
fact, the system is strongly correlated with the environment [Eq.~(\ref%
{Bipartita})].

The system-environment correlations emerges due to a unidirectional
dependence of the system dynamics on the environment transitions [Eq.~(\ref%
{EvolucionRhoTilde})]. In fact, the environment populations do not depend on
the system degrees of freedom [see Eq.~(\ref{Poblaciones})]. These
properties are relevant in the random scheme and imply that $C_{pf}(t,\tau
)|_{\breve{y}}\overset{r}{=}0$ [Eq.~(\ref{CPFforCBE})]. This result is valid
for arbitrary measurement processes, indicating in the operational approach,
the absence of any environment-to-system backflow of information.

\subsection{Environment-to-system backflow of information}

In the previous section we concluded that both approaches differ in the
classification of the dynamics (Markovian vs. non-Markovian), but (due to
different reasons) agree in the absence of any environment-to-system
backflow of information. Here, we show that in general both approaches also
differ in this last aspect. Different mechanisms can be proposed for getting
a revival in the trace distance Eq.~(\ref{TraceDistance}).

\subsubsection{Slow modulation of the stationary environment state}

First, we consider the same model [Eq.~(\ref{EvolucionRhoTilde})], but in
addition it is assumed that the characteristic rates are time dependent, $%
\gamma \rightarrow \gamma (t),$ $\phi \rightarrow \phi (t),$ with%
\begin{equation}
\gamma (t)=\gamma \lbrack 1+b(t)]>0,\ \ \ \ \ \ \phi (t)=\phi \lbrack
1-b(t)]>0.  \label{RateModulated}
\end{equation}%
Here, $b(t)$ is an arbitrary function of time that fulfill the constraint $%
-1<b(t)<1.$ The previous structure is chosen for simplifying the argument
and calculus. Nevertheless, we remark that similar dependences can be
implemented in different experimental situations (see for example Ref.~\cite%
{Udo}). The more relevant aspect is that the assumption (\ref{RateModulated}%
) can be implemented by affecting solely the environmental degrees of
freedom [see Eq.~(\ref{Poblaciones})].

In addition, in Eq.~(\ref{RateModulated}) it is assumed that%
\begin{equation}
\left\vert \frac{d}{dt}b(t)\right\vert \ll \gamma ,\ \ \ \ \left\vert \frac{d%
}{dt}b(t)\right\vert \ll \phi .  \label{slow}
\end{equation}%
Hence, the time dependence of $b(t)$ can be considered slow with respect to
the decay times $(1/\gamma )$ and $(1/\phi ).$ Consequently, the full
dynamics can be described in an adiabatic approximation, where the full
bipartite system in the long time regime $(\gamma t\gg 1,\ \phi t\gg 1)$
rapidly adjust to the instantaneous values of $\gamma (t)$ and $\phi (t).$
In particular, in this regime, the environment populations, from Eq.~(\ref%
{Estacionarias}), can be written as%
\begin{equation}
p_{4}^{\infty }(t)\simeq \frac{\phi }{\gamma +\phi }[1-b(t)],\ \ \ \
p_{k}^{\infty }(t)\simeq \frac{1}{3}\frac{\gamma }{\gamma +\phi }[1+b(t)],
\label{PkSlow}
\end{equation}%
where $k=1,2,3.$ For simplicity, we assumed that $(\gamma -\phi )\ll (\gamma
+\phi ),$ which allows to approximate $\gamma (t)+\phi (t)=(\gamma +\phi
)+b(t)(\gamma -\phi )\simeq (\gamma +\phi ).$

In the long time regime, the \textit{non-operational approach} is
characterized by the value $\lim_{t\rightarrow \infty }w(t)\neq 0$ [see
Eqs.~(\ref{W(t)}) and~(\ref{TraceDistance})]. For time-independent rates
this quantity can be written in terms of the stationary populations $%
\{p_{k}^{\infty }\}_{k=1}^{k=4}$ [Eq.~(\ref{Estacionarias})] as $%
\lim_{t\rightarrow \infty }w(t)=[p_{4}^{\infty
}]^{2}+\sum_{k=1,2,3}[p_{k}^{\infty }]^{2}.$ Given that in the slow
modulation regime [Eq.~(\ref{slow})] these values become time dependent, $%
p_{k}^{\infty }\longrightarrow p_{k}^{\infty }(t)$ [Eq. (\ref{PkSlow})], it
follows that%
\begin{equation}
w(t)\overset{slow}{\simeq }[p_{4}^{\infty
}(t)]^{2}+\sum_{k=1,2,3}[p_{k}^{\infty }(t)]^{2},\ \ \ \ \gamma t\gg 1,\
\phi t\gg 1.
\end{equation}%
Therefore, under the previous hypothesis, the stationary values of the TD in
Fig.~1(a) $[d(t)=\left\vert 4w(t)-1\right\vert /3]$ become proportional to
the arbitrary function $b(t).$ This result implies that one can get \textit{%
arbitrary revivals in the trace distance} [Eq.~(\ref{TraceDistance})] by
choosing different time-dependences of the function $b(t).$ Alternatively,
an \textit{arbitrary environment-to-system backflow of information can be
produced }by changing solely in a slow way the (\textquotedblleft
stationary\textquotedblright ) environment populations. Nevertheless, we
remark that the full dynamics is essentially the same than in the
static-rate case. While one can associate the revivals in the TD to the
system-environment correlations, these correlations have the same origin and
structure than in absence of revivals, that is Fig.~1(a) (static case) and
when $b(t)$ does not lead to revivals.

In the deterministic scheme, the \textit{operational approach} is
characterized by the stationary value \cite{CPF}%
\begin{equation}
\lim_{\substack{ t\rightarrow \infty  \\ \tau \rightarrow \infty }}%
C_{pf}(t,\tau )|_{\breve{y}}\overset{d}{=}\frac{8\gamma (\gamma -3\phi
)^{2}(\gamma +3\phi )}{81(\gamma +\phi )^{4}},
\end{equation}%
which can also be written in terms of $\{p_{k}^{\infty }\}_{k=1}^{k=4}$
[Eq.~(\ref{Estacionarias})]. Thus, under the same conditions that guarantee
the slow modulation regime [Eqs.~(\ref{slow}) and (\ref{PkSlow})], the
stationary values of $C_{pf}(t,t)|_{\breve{y}}$ plotted in Fig.~1(b) also
become proportional to the function $b(t).$ Nevertheless, in this approach
this property does not implies the presence of any backflow of information.
In fact, given that the environment state does not depend at all on the
system degrees of freedom, even in the slow modulation regime, it follows
that $C_{pf}(t,\tau )|_{\breve{y}}\overset{r}{=}0$ [Eq.~(\ref{CPFforCBE})].
In this way, it is clear that both the non-operational and operational
approaches also strongly disagree in this aspect.

\subsubsection{Quantum coherent contributions in the environment dynamics}

The system-environment dynamics associated to the depolarizing channel [Eq.~(%
\ref{EvolucionRhoTilde})] can alternatively be represented through a
Lindblad equation. In fact, the evolution of the bipartite state $\rho
_{t}^{se}$ can be written as%
\begin{eqnarray}
\frac{d\rho _{t}^{se}}{dt} &=&+\frac{\gamma }{3}\sum_{k=1,2,3}(B_{k}\sigma
_{k}[\rho _{t}^{se}]\sigma _{k}B_{k}^{\dagger }-\frac{1}{2}\{B_{k}^{\dagger
}B_{k},\rho _{t}^{se}\}_{+})  \notag \\
&&+\phi \sum_{k=1,2,3}(B_{k}^{\dagger }\sigma _{k}[\rho _{t}^{se}]\sigma
_{k}B_{k}-\frac{1}{2}\{B_{k}B_{k}^{\dagger },\rho _{t}^{se}\}_{+})  \notag \\
&&-i[H_{e},\rho _{t}^{se}],  \label{BipartitoLindblad}
\end{eqnarray}%
where the bath operators are $B_{k}\equiv |k\rangle \langle 4|,$ $k=1,2,3.$
As before, $\{|1\rangle ,|2\rangle ,|3\rangle ,|4\rangle \}$ are the
environment base. Defining the states $\tilde{\rho}_{k}\equiv \langle k|\rho
_{t}^{se}|k\rangle ,$ it is simple to check that the first two lines of the
previous Lindblad dynamics recover the time evolution introduced in Eq.~(\ref%
{EvolucionRhoTilde}).

From Eq.~(\ref{BipartitoLindblad}) it is simple to check that the bath state 
$(\rho _{t}^{e}=\mathrm{Tr}_{s}[\rho _{t}^{se}])$ obeys a Lindblad equation
that, even with the extra contribution $-i[H_{e},\rho _{t}^{se}],$ is
independent of the system degrees of freedom. Thus, the environment is still
a casual bystander one [see Eqs.~(\ref{GeneralSU}) and (\ref{BathLindblad}%
)]. In order to obtain a (system) depolarizing channel [Eq.~(\ref{DepoRho})]
the symmetry between the bath states $\{|1\rangle ,|2\rangle ,|3\rangle \}$
must be granted. For example, the Hamiltonian%
\begin{equation}
H_{e}=\frac{\Omega }{2}\sum_{k=1,2,3}(|k\rangle \langle 4|+|4\rangle \langle
k|),
\end{equation}%
fulfill this property. 
\begin{figure}[tbp]
\includegraphics[bb=22 375 550 595,angle=0,width=8.5cm]{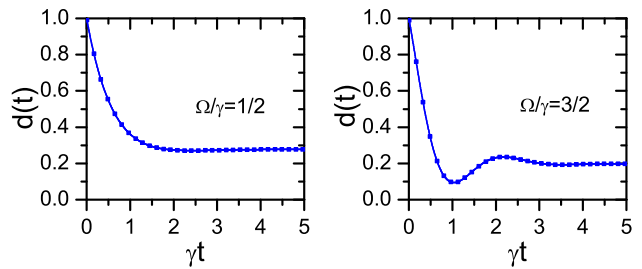}
\caption{Decay of the trace distance $d(t)$ [Eqs.~(\protect\ref%
{TraceDistance}) and (\protect\ref{w_Unitario})] corresponding to the model (%
\protect\ref{BipartitoLindblad}) with $\protect\phi =\protect\gamma $ for
different values of the Hamiltonian frequency $\Omega /\protect\gamma .$}
\end{figure}

In consistence with the solution defined by Eqs.~(\ref{Bipartita}) and (\ref%
{RhoTildeK}), here the bipartite state is written as%
\begin{equation}
\rho _{t}^{se}=\sum_{k=1,2,3,4}(\sigma _{k}\rho _{0}^{s}\sigma _{k})\otimes
\varrho _{t}^{k},
\end{equation}%
where $\{\varrho _{t}^{k}\}$ are states in the environment Hilbert space. In
order to obtain analytical treatable solutions we assume the bipartite
initial condition%
\begin{equation}
\rho _{0}^{se}=\rho _{0}^{s}\otimes \rho _{0}^{e}=\rho _{0}^{s}\otimes
|4\rangle \langle 4|.  \label{CIN4}
\end{equation}%
Under this assumption $(\rho _{0}^{e}=|4\rangle \langle 4|),$ given that the
underlying system stochastic dynamics associated to Eq.~(\ref%
{BipartitoLindblad}) is the same than in the incoherent case [Eq.~(\ref%
{EvolucionRhoTilde})], it follows that the system state goes back to the
initial condition $\rho _{0}^{s}$ whenever the environment goes back to the
state $|4\rangle .$ This property straightforwardly follows from $\sigma
_{k}^{2}=\mathrm{I}.$ Therefore, under the assumption~(\ref{CIN4}), here the
depolarizing map Eq.~(\ref{DepoRho}) is defined with the function%
\begin{equation}
w(t)=\mathrm{Tr}_{e}[\varrho _{t}^{4}]=\langle 4|\rho _{t}^{e}|4\rangle ,
\label{w_Unitario}
\end{equation}%
where $\rho _{t}^{e}$ is the density matrix of the environment. Consistently 
$[1-w(t)]/3=\mathrm{Tr}_{e}[\varrho _{t}^{k}]=\langle k|\rho
_{t}^{e}|k\rangle ,$ with $k=1,$ $2,$ $3.$ In consequence, the decay of the
trace distance is proportional to the bath population $\langle 4|\rho
_{t}^{e}|4\rangle .$ Its explicit analytical expression is rather complex
and non-informative \cite{p4}.

In this alternative situation, it is clear that $H_{e}$ induces intrinsic
quantum coherent oscillations in the environment dynamics, which in turn may
lead to oscillations in the trace distance [Eq.~(\ref{TraceDistance})]. In
Fig.~2 we plot the TD decay $d(t)=\left\vert 4w(t)-1\right\vert /3$ taking $%
\phi =\gamma $ and for different values of $\Omega /\gamma .$ When $\Omega
/\gamma <1,$ a monotonous decay is observed. Nevertheless, for $\Omega
/\gamma >1$ \textit{revivals} in the TD are observed.

The CPF correlation in the deterministic scheme cannot be calculated in an
analytical way. Nevertheless, given that the system dynamics is still
controlled by the environment (self) transitions, it follows that $%
C_{pf}(t,\tau )|_{\breve{y}}\overset{d}{\neq }0.$ Thus, the dynamics becomes
non-Markovian in both approaches $(\Omega /\gamma >1).$ Nevertheless, given
that the environment is a casual bystander one, in the random scheme it is
valid that $C_{pf}(t,\tau )|_{\breve{y}}\overset{r}{=}0$ [Eq.~(\ref%
{CPFforCBE})] for any value of $\Omega /\gamma .$ Consequently, in the same
way as in the previous model [Eq.~(\ref{RateModulated})] the non-operational
and operational approaches gives different results about the presence of
environment-to-system backflows of information.

\section{Summary and conclusions}

The interpretation of quantum memory effects in terms of an
environment-to-system backflow of information is still under a vivid debate.
In this contribution we presented a partial view of this problem by
comparing how this concept is introduced an interpreted in non-operational
and operational approaches to quantum non-Markovianity.

Our main contribution is a comparison between both formalisms for different
environment models. We considered casual bystander environments, which are
characterized by a density matrix that does not depend on the system degrees
of freedom. This class cover classical statistical mixtures of Markovian
dynamics [Eq.~(\ref{Superposition})], interaction with stochastic classical
degrees of freedom [Eq.~(\ref{BipartitoClassical})], and also environmental
quantum degrees of freedom [Eq.~(\ref{BipartitoCBE})]. In addition, we
considered unitary system-environment models [Eq.~(\ref{Htotal})].

As a non-operational approach we used the TD between two system states with
different initial conditions. This formalism is characterized by the bound
Eq.~(\ref{General}). We have argued that, in general, it is not possible to
predict if for a given model the TD presents or not revivals in its time
behavior. This property is valid for all environmental models. In the case
of casual bystander ones, the previous feature represent an obstacle for
giving a consistent physical interpretation of any environment-to-system
backflow of information defined as revivals in the TD [Eq.~(\ref{Revivals}%
)]. In fact, for these dynamics the system-environment correlations emerges
due to a unidirectional dependence of the system dynamics in the state of
the environment and its transitions. In particular, for stationary
environments it is not possible to know when the system-environment
correlations lead to the presence or absence of backflows of information.
The possibility of getting monotonous decay behaviors of the TD for unitary
interaction models also represents an undesirable property because in
general the environment state is modified by its interaction with the system.

As an operational approach we used a CPF correlation [Eq.~(\ref{CPFCorre})],
which is defined by three consecutive system measurement processes. Both a
deterministic and random schemes were considered [with associated joint
probabilities Eqs.~(\ref{Joint_d}) and~(\ref{Joint_r})]. In the case of
casual bystander environments, the CPF correlation in the deterministic
scheme does not vanish, while in the random scheme it vanishes identically
for any chosen measurement observables [Eq.~(\ref{CPFforCBE})]. Thus, in
this approach, any casual bystander environment leads to a non-Markovian
system dynamics but not any bidirectional information flow is detected. In
the case of Hamiltonian models, in general, in both schemes the CPF
correlation does not vanish indicating a non-Markovian system dynamics and
the presence of bidirectional information flows [Eq.~(\ref{CPFUnitary})]. An
undesirable exception to this last property emerges when the system dynamics
can equivalently be represented by a random unitary map [Eqs.~(\ref%
{DiagonalCondition}) and (\ref{SystemRandomUnitary})].

 As an specific example we considered a system coupled to an
environment able to induce a depolarizing dynamics [Eqs.~(\ref%
{EvolucionRhoTilde}), (\ref{RateModulated}) and (\ref{BipartitoLindblad})].
We have found that both approaches differ in the Markovian and non-Markovian
regimes as well in the presence or absence of environment-to-system
backflows of information.

In general, both operational and non-operational approaches to quantum
non-Markovianity provide necessary and complementary points of view for
defining and understanding memory effects in open quantum systems. The
present results shed light on some conceptual differences and properties of
these approaches. They may be useful for extending the application of these
formalisms for the understanding of memory effects induced by structured or
spatially extended environments.

\section*{Acknowledgments}

This paper was supported by Consejo Nacional de Investigaciones Cient\'{\i}%
ficas y T\'{e}cnicas (CONICET), Argentina.


\begin{thebibliography}{99}
\bibitem{vanKampen} N. G. van Kampen, \textit{Stochastic Processes in
Physics and Chemistry}, (North-Holland, Amsterdam, 1992).

\bibitem{breuerbook} H. P. Breuer and F. Petruccione, \textit{The theory of
open quantum systems}, (Oxford University press, 2002).

\bibitem{vega} I. de Vega and D. Alonso, Dynamics of non-Markovian open
quantum systems, Rev. Mod. Phys. \textbf{89}, 015001 (2017).

\bibitem{wiseman} L. Li, M. J. W. Hall, and H. M. Wiseman, Concepts of
quantum non-Markovianity: A hierarchy, Phys. Rep. \textbf{759}, 1 (2018).

\bibitem{BreuerReview} H. P. Breuer, E. M. Laine, J. Piilo, and V. Vacchini,
Colloquium: Non-Markovian dynamics in open quantum systems, Rev. Mod. Phys. 
\textbf{88}, 021002 (2016); H. P. Breuer, Foundations and measures of
quantum non-Markovianity, J. Phys. B \textbf{45}, 154001 (2012).

\bibitem{plenioReview} A. Rivas, S. F. Huelga, and M. B. Plenio, Quantum
non-Markovianity: characterization, quantification and detection, Rep. Prog.
Phys. \textbf{77}, 094001 (2014).

\bibitem{divisible} M. M.Wolf, J. Eisert, T. S. Cubitt, and J. I. Cirac,
Assessing Non-Markovian Quantum Dynamics, Phys. Rev. Lett. \textbf{101},
150402 (2008); A. Rivas, S. F. Huelga, and M. B. Plenio, Entanglement and
Non-Markovianity of Quantum Evolutions, Phys. Rev. Lett. \textbf{105},
050403 (2010). 

\bibitem{BreuerFirst} H. P. Breuer, E. M. Laine, and J. Piilo, Measure for
the Degree of Non-Markovian Behavior of Quantum Processes in Open Systems,
Phys. Rev. Lett. \textbf{103}, 210401 (2009); E. M. Laine, J. Piilo, and H.
P. Breuer, Measure for the non-Markovianity of quantum processes, Phys. Rev.
A \textbf{81}, 062115 (2010).

\bibitem{EnergyBackFLow} G. Guarnieri, C. Uchiyama, and B. Vacchini, Energy
backflow and non-Markovian dynamics, Phys. Rev. A \textbf{93}, 012118 (2016).

\bibitem{Energy} G. Guarnieri, J. Nokkala, R. Schmidt, S. Maniscalco, and B.
Vacchini, Energy backflow in strongly coupled non-Markovian
continuous-variable systems, Phys. Rev. A \textbf{94}, 062101 (2016).

\bibitem{HeatBackFLow} R. Schmidt, S. Maniscalco, and T. Ala-Nissila, Heat
flux and information backflow in cold environments, Phys. Rev. A \textbf{94}%
, 010101(R) (2016).

\bibitem{megier} N. Megier, D. Chru\'{s}ci\'{n}ski, J. Piilo, and W. T.
Strunz, Eternal non-Markovianity: from random unitary to Markov chain
realisations, Sci. Rep. \textbf{7}, 6379 (2017).

\bibitem{maximal} A. A. Budini, Maximally non-Markovian quantum dynamics
without environment-to-system backflow of information, Phys. Rev. A \textbf{%
97}, 052133 (2018).

\bibitem{petruccione} F. A. Wudarski and F. Petruccione, Exchange of
information between system and environment: Facts and myths, Euro Phys.
Lett. \textbf{113}, 50001 (2016).

\bibitem{amato} H. P. Breuer, G. Amato, and B. Vacchini, Mixing-induced
quantum non-Markovianity and information flow, New J. Phys. \textbf{20},
043007 (2018).

\bibitem{santis} D. De Santis and M. Johansson, Equivalence between
non-Markovian dynamics and correlation backflows, New J. Phys. \textbf{22},
093034 (2020).

\bibitem{acin} D. De Santis, M. Johansson, B. Bylicka, N. K. Bernardes, and
A. Ac\'{\i}n, Witnessing non-Markovian dynamics through correlations, Phys.
Rev. A \textbf{102}, 012214 (2020).

\bibitem{horo} M. Banacki, M. Marciniak, K. Horodecki, and P. Horodecki,
Information backflow may not indicate quantum memory, arXiv:2008.12638.

\bibitem{EntroBack} N. Megier, A. Smirne, and B. Vacchini, Entropic Bounds
on Information Backflow, Phys. Rev. Lett. \textbf{127}, 030401 (2021); S.
Campbell, M. Popovic, D. Tamascelli, and B. Vacchini, Precursors of
non-Markovianity, New J. Phys. \textbf{21}, 053036, (2019).

\bibitem{modi} F. A. Pollock, C. Rodr\'{\i}guez-Rosario, T. Frauenheim, M.
Paternostro, and K. Modi, Operational Markov Condition for Quantum
Processes, Phys. Rev. Lett. \textbf{120}, 040405 (2018); F. A. Pollock, C.
Rodr\'{\i}guez-Rosario, T. Frauenheim, M. Paternostro, and K. Modi,
Non-Markovian quantum processes: Complete framework and efficient
characterization, Phys. Rev. A \textbf{97}, 012127 (2018).

\bibitem{budiniCPF} A. A. Budini, Quantum Non-Markovian Processes Break
Conditional Past-Future Independence, Phys. Rev. Lett. \textbf{121}, 240401
(2018); A. A. Budini, Conditional past-future correlation induced by
non-Markovian dephasing reservoirs, Phys. Rev. A \textbf{99}, 052125 (2019).

\bibitem{pollock} P. Taranto, F. A. Pollock, S. Milz, M. Tomamichel, and K.
Modi, Quantum Markov Order, Phys. Rev. Lett. \textbf{122}, 140401 (2019); P.
Taranto, S. Milz, F. A. Pollock, and K. Modi, Structure of quantum
stochastic processes with finite Markov order, Phys. Rev. A \textbf{99},
042108 (2019).

\bibitem{pollockInfluence} M. R. J\o rgensen and F. A. Pollock, Exploiting
the Causal Tensor Network Structure of Quantum Processes to Efficiently
Simulate Non-Markovian Path Integrals, Phys. Rev. Lett. \textbf{123}, 240602
(2019).

\bibitem{bonifacio} M. Bonifacio and A. A. Budini, Perturbation theory for
operational quantum non-Markovianity, Phys. Rev. A \textbf{102}, 022216
(2020).

\bibitem{han} L. Han, J. Zou, H. Li, and B. Shao, Non-Markovianity of A
Central Spin Interacting with a Lipkin--Meshkov--Glick Bath via a
Conditional Past--Future Correlation, Entropy \textbf{22}, 895 (2020).

\bibitem{ban} M. Ban, Operational non-Markovianity in a statistical mixture
of two environments, Phys. Lett. A \textbf{397}, 127246 (2021).

\bibitem{rio} T. de Lima Silva, S. P. Walborn, M. F. Santos, G. H. Aguilar,
and A. A. Budini, Detection of quantum non-Markovianity close to the
Born-Markov approximation, Phys. Rev. A \textbf{101}, 042120 (2020).

\bibitem{goan} Y. -Y. Hsieh, Z. -Y. Su, and H. -S. Goan, Non-Markovianity,
information backflow, and system-environment correlation for
open-quantum-system processes, Phys. Rev. A \textbf{100}, 012120 (2019).

\bibitem{BIF} A. A. Budini, Detection of bidirectional system-environment
information exchanges, Phys. Rev. A \textbf{103}, 012221 (2021).

\bibitem{casual} A. A. Budini, Quantum non-Markovian \textquotedblleft
casual bystander\textquotedblright\ environments, Phys. Rev. A \textbf{104},
062216 (2021).

\bibitem{alicki} R. Alicki and K. Lendi, Quantum Dynamical Semigroups and
Applications, Lect. Notes Phys. \textbf{717} (Springer, Berlin Heidelberg,
2007).

\bibitem{nielsen} M. A. Nielsen and I. L. Chuang, \textit{Quantum
Computation and Quantum Information} (Cambridge University Press, Cambridge,
2000).

\bibitem{poland} D. Chruscinski and F. A. Wudarski, Non-Markovian random
unitary qubit dynamics, Phys. Lett. A \textbf{377}, 1425 (2013);
Non-Markovianity degree for random unitary evolution, Phys. Rev. A \textbf{91%
}, 012104 (2015); F. A. Wudarski, P. Nalezyty, G. Sarbicki, and D.
Chruscinski, Admissible memory kernels for random unitary qubit evolution, 
\textit{ibid.} \textbf{91}, 042105 (2015); F. A. Wudarski and D.
Chruscinski, Markovian semigroup from non-Markovian evolutions, \textit{ibid.%
} \textbf{93}, 042120 (2016); K. Siudzinska and D. Chruscinski, Memory
kernel approach to generalized Pauli channels: Markovian, semi-Markov, and
beyond, ibid. \textbf{96}, 022129 (2017).

\bibitem{lidar} C. Sutherland, T. A. Brun, and D. A. Lidar, Non-Markovianity
of the post-Markovian master equation, Phys. Rev. A \textbf{98}, 042119
(2018); A. Shabani and D. A. Lidar, Completely positive post-Markovian
master equation via a measurement approach, \textit{ibid.} \textbf{71},
020101(R) (2005); A. A. Budini, Post-Markovian quantum master equations from
classical environment fluctuations, Phys. Rev. E \textbf{89}, 012147 (2014).

\bibitem{collision} B. Vacchini, Non-Markovian master equations from
piecewise dynamics, Phys. Rev. A \textbf{87}, 030101(R) (2013); A. A.
Budini, Embedding non-Markovian quantum collisional models into bipartite
Markovian dynamics, Phys. Rev. A \textbf{88}, 032115 (2013); A. A. Budini
and P. Grigolini, Non-Markovian nonstationary completely positive
open-quantum-system dynamics, \textit{ibid.} \textbf{80}, 022103 (2009).

\bibitem{entanglement} R. Horodecki, P. Horodecki, M. Horodecki, and K.
Horodecki, Quantum entanglement, Rev. Mod. Phys. \textbf{81}, 865 (2009).

\bibitem{katarzyna}  K. Roszak and \L ukasz Cywi\'{n}ski,
Characterization and measurement of qubit-environment-entanglement
generation during pure dephasing, Phys. Rev. A 92, 032310 (2015); K. Roszak
and \L ukasz Cywi\'{n}ski, Equivalence of qubit-environment entanglement and
discord generation via pure dephasing interactions and the resulting
consequences, Phys. Rev. A 97, 012306 (2018); K. Roszak, Criteria for
system-environment entanglement generation for systems of any size in
pure-dephasing evolutions, Phys. Rev. A 98, 052344 (2018).

\bibitem{nori} H.-B. Chen, C. Gneiting, P.-Y. Lo, Y.-N. Chen, and F. Nori,
Simulating Open Quantum Systems with Hamiltonian Ensembles and the
Nonclassicality of the Dynamics, Phys. Rev. Lett. \textbf{120}, 030403
(2018).

\bibitem{multi} A. A. Budini and J. P. Garrahan, Solvable
class of non-Markovian quantum multipartite dynamics, Phys. Rev. A \textbf{%
104}, 032206 (2021).

\bibitem{CPF} $C_{pf}(t,\tau )|_{\breve{y}}\overset{d}{=}\frac{%
8}{81(\gamma +\phi )^{4}}e^{-2t\gamma -\tau \gamma -3t\phi -2\tau \phi
}\gamma $ $((-2e^{(t+\tau )(\gamma +2\phi )}(\gamma -3\phi )^{2}\phi
-2e^{2t\gamma +3t\phi +\tau \phi }(\gamma -3\phi )^{2}\phi -e^{2t\gamma
+\tau \gamma +3t\phi +\tau \phi }(\gamma -3\phi )^{2}(\gamma +\phi
)-e^{2t(\gamma +\phi )+\tau (\gamma +2\phi )}(\gamma -3\phi )^{2}(\gamma
+\phi )-16e^{\tau \phi +2t(\gamma +\phi )}\gamma \phi (\gamma +\phi
)-16e^{\tau (\gamma +\phi )+t(\gamma +2\phi )}\gamma \phi (\gamma +\phi
)+e^{2t\gamma +\tau \gamma +3t\phi +2\tau \phi }(\gamma -3\phi )^{2}(\gamma
+3\phi )+e^{(2t+\tau )(\gamma +\phi )}(\gamma +\phi )^{2}(\gamma +9\phi
)+2e^{t\gamma +2t\phi +\tau \phi }\phi \left( 9\gamma ^{2}+2\gamma \phi
+9\phi ^{2}\right) ).$ 

\bibitem{Udo} S. Schuler, T. Speck, C. Tietz, J. Wrachtrup,
and U. Seifert, Experimental Test of the Fluctuation Theorem for a Driven
Two-Level System with Time-Dependent Rates, Phys. Rev. Lett. \textbf{94},
180602 (2005). 

\bibitem{p4} In the Laplace domain, $f(u)=\int_{0}^{\infty }dte^{-ut}f(t),$
it reads $\langle 4|\rho _{u}^{e}|4\rangle =A(u)/B(u),$ where $A(u)=\gamma
(u+\phi )(2u+\gamma +\phi )+(3(u+\phi )+2\gamma )\Omega ^{2}$ and $%
B(u)=3u(u+\phi )(u+\gamma +\phi )(2u+\gamma +\phi )+6u(3(u+\phi )+\gamma
)\Omega ^{2}.$
\end{thebibliography}
\end{document}